\documentclass[journal,a4paper]{IEEEtran}
\ifCLASSINFOpdf
\else
\fi
\usepackage{amsmath, amssymb, bm, cite, epsfig, psfrag}
\usepackage{epstopdf}
\usepackage{graphicx}
\usepackage{algorithm,algpseudocode}
\usepackage{array}
\usepackage[margin=.81in]{geometry}
\usepackage[margin=10pt,font=small]{caption}
\usepackage{bbm}
\usepackage{multirow}
\usepackage[usenames,dvipsnames]{xcolor}

\usepackage{subfigure}
\usepackage{etoolbox}
\usepackage{pbox}
\usepackage[width=0.48\textwidth]{caption}
\usepackage{colortbl}

\newtoggle{conference}
\togglefalse{conference} 
\graphicspath{{figures/}}

\def\beq{\begin{equation}}
\def\eeq{\end{equation}}
\def\beqa{\begin{eqnarray}}
\def\eeqa{\end{eqnarray}}
\def\beqan{\begin{eqnarray*}}
\def\eeqan{\end{eqnarray*}}

\def\tm1{t\! - \! 1}
\def\tp1{t\! + \! 1}

\usepackage{lipsum}
\usepackage{fancyhdr}
\usepackage{datetime}

\usepackage{tikz}
\usetikzlibrary{calc}

\begin{document}
%
\title{Local Multipath Model Parameters for Generating 5G Millimeter-Wave 3GPP-like Channel Impulse Response}

\author{
\IEEEauthorblockN{Mathew K. Samimi, Theodore S. Rappaport}\\
\IEEEauthorblockA{NYU WIRELESS, NYU Tandon School of Engineering}\\
\IEEEauthorblockA{mks@nyu.edu, tsr@nyu.edu}

\thanks{The authors wish to thank the NYU WIRELESS Industrial Affiliates for their support. This work is supported by National Science Foundation (NSF) Grants (1302336, 1320472, and 1555332).}

}



\maketitle

\begin{tikzpicture} [remember picture, overlay]
\node at ($(current page.north) + (0,-0.25in)$) {M. K. Samimi, T. S. Rappaport, ``Local Multipath Model Parameters for Generating 5G Millimeter-Wave};
\node at ($(current page.north) + (0,-0.4in)$) { 3GPP-like Channel Impulse Response,'' \textit{in the 10\textsuperscript{th} European Conference on Antennas and Propagation}};
\node at ($(current page.north) + (0,-0.6in)$) { \textit{ (EuCAP'2016)}, April 2016.};
\end{tikzpicture}

\begin{abstract}
This paper presents 28 GHz and 73 GHz empirically-derived large-scale and small-scale channel model parameters that characterize average temporal and angular properties of multipaths. Omnidirectional azimuth scans at both the transmitter and receiver used high gain directional antennas, from which global 3GPP modeling parameters for the mean global azimuth and zenith spreads of arrival were found to be 22$^{\circ}$ and 6.2$^{\circ}$ at 28 GHz, and 37.1$^{\circ}$ and 3.8$^{\circ}$ at 73 GHz, respectively, in non-line of sight (NLOS). Small-scale spatial measurements at 28 GHz reveal a mean cross-polar ratio for individual multipath components of 29.7 dB and 16.7 dB in line of sight and NLOS, respectively. Small-scale parameters extracted using the KPowerMeans algorithm yielded on average 5.3 and 4.6 clusters at 28 GHz and 73 GHz, respectively, in NLOS. The time cluster - spatial lobe (TCSL) modeling approach uses an alternative physically-based binning procedure and recreates 3GPP model parameters to generate channel impulse responses, as well as new parameters like the RMS lobe angular spreads useful in quantifying millimeter-wave directionality. The TCSL algorithm faithfully reproduces first- and second-order statistics of measured millimeter-wave channels. 
\end{abstract}


%
\IEEEpeerreviewmaketitle

\vspace{7pt}
\section{Introduction}
The channel impulse response of a radio-propagation channel is composed of multipath components, whose local average temporal and angular characteristics can be obtained from large-scale and small-scale parameters. Large-scale parameters usually denote the azimuth spread (AS), the root-mean-square (RMS) delay spread, the shadow fading (SF), and the Rician $K$-factor~\cite{3GPP:1,WinnerII}, but must also be extended to include zenith (i.e., elevation) spreads, as directionality in azimuth and elevation is expected to drive future mmWave systems through multi-antenna arrays~\cite{Rap15_3}. While large-scale parameters for channel impulse responses describe typical (average) local properties of a radio-channel (over a local area of tens of wavelengths), it is not to be confused with \textit{large-scale} path loss which accounts for signal level fluctuations over several \textit{thousands} of wavelengths resulting from large obstructions in the wireless channel (such as buildings). The 3GPP and WINNER II spatial channel models (SCMs) model the power delay profile (PDP) by utilizing the large-scale parameters as first- and second-order inputs to statistical distributions that govern the statistics of small-scale parameters of multipaths. Small-scale parameters refer to the properties of a single multipath component, such as path delay, and angles of arrival and departure. By way of contrast, the COST 2100 model does not explicitly generate large-scale parameters, but instead statistically generates an environment that is independent of mobile station (MS) location, and synthesizes the large-scale parameters based on cluster scattering~\cite{Liu12}. 5G millimeter-wave (mmWave) channel models do not yet exist in the 3GPP and ITU standards, driving the need for channel model standards for future mmWave technologies, with large bandwidths (bit rates) and directional antenna gains (multi-element antenna arrays)~\cite{Rap15_3}.

The cross-correlation of large-scale parameters enhances the spatial consistency and accuracy of system-level simulations in which many users within close proximity must experience realistic correlated channels as observed in~\cite{Algans02,3GPP:1,WinnerII}. In the 3GPP model, the shadow fading values on the omnidirectional received powers for two users separated in space at the same time instant are generated from two Gaussian random variables (in dB) with a correlation coefficient of 0.5~\cite{3GPP:1}, based on observations in~\cite{Algans02}. The WINNER II urban microcellular (UMi) models generate the path delays from an exponential distribution whose mean is a function of the RMS delay spread (see Eq.~(4.1) in~\cite{WinnerII}), and generate path angles using a wrapped Gaussian function whose standard deviation is a function of second-order measured angular spreads (see Eq.~(4.8) in~\cite{WinnerII}). Both large-scale and small-scale path parameters are thus critical in generating wideband channel impulse responses that recreate the statistics of a large ensemble of collected measurements. The large-scale and small-scale parameters of 28 GHz and 73 GHz mmWave channels are provided, obtained from measurements carried out in 2012, 2013~\cite{Rap13:2,Rap15_3}, and 2015~\cite{Samimi16} in New York City.

\vspace{7pt}
\section{Measurement Descriptions}

\begin{figure}[t]
    \centering
 \includegraphics[width=3.4in]{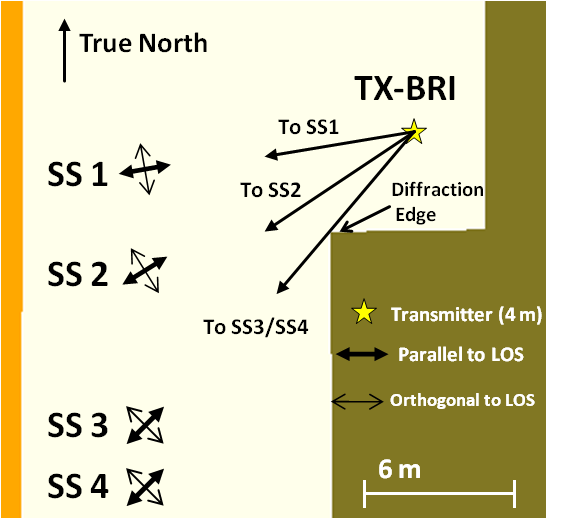}
    \caption{Map of the environment in which the 28 GHz small-scale fading measurements were collected. One TX on Bridge (BRI) Street and four RX locations were selected in LOS (SS 1 and SS 2), LOS-to-NLOS (SS 3), and NLOS (SS 4) scenarios. The two orthogonal arrows represent the two axes of a cross, over which a PDP was sampled every 5.35 mm to emulate a virtual array of directional antenna elements.}
    \label{fig:map}
\end{figure}

Two outdoor propagation measurement campaigns were performed at 74 RX and 36 RX locations at 28 GHz and 73 GHz~\cite{Rap13:2,Rap15_3}, respectively, using a 400 megachips-per-second broadband sliding correlator channel sounder, and a pair of 24.5 dBi and 27 dBi directional antennas that provided over 12,000 directional PDPs to study AOD and AOA statistics of the mmWave channel~\cite{Rap15_3,Samimi15_3}. Additional 28 GHz small-scale fading measurements investigated the statistics of cross-polarization ratio (XPR) of individual multipath component amplitudes over an outdoor local area at one TX and four RX locations in line of sight (LOS), LOS-to-NLOS, and NLOS environments for distances ranging from 8 m to 12.9 m, using a pair of 15 dBi (28.8$^{\circ}$ and 30$^{\circ}$ half-power beamwidths in azimuth and elevation, respectively) gain horn antennas at the TX and RX, for vertical-to-vertical and vertical-to-horizontal polarization configurations.  The measurements emulated a single-input multiple-output (SIMO) realistic base-to-mobile communication link, with the TX and RX placed 4 m and 1.4 m above ground, respectively. The PDPs were spatially sampled every $\lambda / 2=$ 5.35 mm on a linear track by placing the RX antenna over two axes of a cross to emulate a virtual array (shown as two orthogonal arrows in Fig.~\ref{fig:map}), and capturing a PDP at static linear track position. Fig.~\ref{fig:map} shows a map of the environment where the small-scale track measurements were collected~\cite{Samimi16}.

\vspace{7pt}
\section{Large-Scale Parameters}

The global azimuth spread quantifies angular dispersion in the omnidirectional azimuth plane, computed as in~(\ref{eq1})~\cite{3GPP:1}:
\begin{equation}\label{eq1}
\sigma_{\theta} = \min_{\Delta}\sqrt{\overline{\theta(\Delta)^2}-\overline{\theta(\Delta)}^2}
\end{equation}
\noindent where,
\begin{equation}\label{eq2}
\overline{\theta(\Delta)} = \frac{\sum_i P(\theta_i) \times \big( (\Delta + \theta_i) (\text{mod } 2\pi) \big)}{\sum_i P(\theta_i)}
\end{equation}

\begin{equation}\label{eq3}
\overline{\theta(\Delta)^2} = \frac{\sum_i P(\theta_i) \times \big( (\Delta + \theta_i) (\text{mod } 2\pi) \big)^2}{\sum_i P(\theta_i)}
\end{equation}

\noindent where $P(\theta)$ is the power azimuth spectrum, $\theta_i$  is the azimuth AOA of path $i$, and $\Delta \in [0,2\pi)$. Since azimuth spreads are circular from 0 to $2\pi$, taking the minimum value with respect to the dummy variable $\Delta$ removes the $2\pi$ ambiguity. The global zenith spread can be computed from~(\ref{eq1}),~(\ref{eq2}), and~(\ref{eq3}), by replacing the azimuth angles $\theta_i$ with the elevation angles $\phi_i$. 

The global (i.e., omnidirectional) RMS delay spread is a measure of channel temporal dispersion, and is defined as the second central moment of a PDP~\cite{Rap15_3}. The Rican $K$-factor specifies the ratio of the strongest multipath power $P_{max}$ to the sum of powers of the other weaker multipaths, as in~(\ref{e22}):
\begin{equation}\label{e22}
K = \frac{P_{max}}{P_{tot}-P_{max}}
\end{equation}
\noindent where $P_{tot}$ is the total received power from all multipath components (e.g. area under the PDP curve). The $K$-factors were computed to determine whether strong multipath components exist in LOS and NLOS environments. 

The global zenith spreads of arrival (ZSA) were observed to exhibit a dependence on transmitter-receiver (T-R) separation $d$, motivating a local mean model. The local means of the $\log_{10}(ZSA)$ were modelled following the 3GPP approach~\cite{3GPP:2}: 
\begin{equation}\label{e6}
\mu_{ZSA}(d) = \max(a \times d+b, c)
\end{equation}

\noindent where the coefficients $a$, $b$, and $c$ are shown in Table~\ref{tbl:T1}, obtained using the minimum mean square error (MMSE) method, by minimizing the error between~(\ref{e6}) and the local mean of $\log_{10}(ZSA)$ values. The ZSDs were not computed because of the limited TX elevation diversity in the measurements~\cite{Rap15_3}.

The global parameters in Table~\ref{tbl:T1} and cross-correlation coefficients in Table~\ref{tbl:T2}, and small-scale parameters in Table~\ref{tbl:T3} were computed from a subset of previously used data presented in~\cite{Samimi15_3,Rap15_3}, where all RX locations selected in this analysis had both an available omnidirectional PDP and TX/RX power angular spectra. The computed LSPs were obtained from 3 and 5 locations in LOS, and 13 and 19 locations in NLOS~\cite{MacCartney15}, at 28 GHz and 73 GHz, respectively. 

\begin{table}[t]
\centering
\caption{Table summarizing large-scale parameters obtained from 28 GHz and 73 GHz channel measurements.}
\resizebox{8.5cm}{!} {
\begin{tabular}{|c|c|c|c|c|}
\hline

\multicolumn{2}{|c|}{\multirow{2}{*}{Scenario}} & \multicolumn{3}{c|}{Urban Microcell (UMi)} \tabularnewline \cline{3-5}
\multicolumn{2}{|c|}{}	& LOS		& \multicolumn{2}{c|}{NLOS}				\tabularnewline \cline{1-5}

\multicolumn{2}{|c|}{Parameters}					& 28-73 GHz Combined	& 28 GHz 	& 73 GHz \tabularnewline \hline

\multirow{2}{*}{DS}	&	$\text{med}$ (ns)			&17.5				& 29.9	& 44.8 \tabularnewline \cline{2-5}

				&	$\mu$ (ns)				& 26.6			& 42.1	& 45.9 \tabularnewline \hline

\multirow{2}{*}{log(DS)}	&	$\mu $				& -7.71			& -7.64	& -7.53 \tabularnewline \cline{2-5}

				&	$\sigma$				& 0.34			& 0.50	& 0.51 \tabularnewline \hline

\multirow{2}{*}{ASD}	&	$\text{med}$ $(^{\circ})$		& 18.5			& 30.9	&26.0 \tabularnewline \cline{2-5}

				&	$\mu$ $(^{\circ})$ 			& 32.3			& 33.7	&28.96 \tabularnewline \hline	

\multirow{2}{*}{log(ASD)}&	$\mu$ 				& 1.28			& 1.38	&1.34 \tabularnewline \cline{2-5}

				&	$\sigma$				& 0.50			& 0.41	& 0.39 \tabularnewline \hline

\multirow{2}{*}{ASA}	& $\text{med}$ $(^{\circ})$ 		& 50.9 			& 22.0 	& 37.1 \tabularnewline \cline{2-5}
				& $\mu$ $(^{\circ})$ 			& 56.9			& 22.0	& 37.1 \tabularnewline \hline

\multirow{2}{*}{log(ASA)}& $\mu$ 					& 1.69			& 1.39	& 1.50 \tabularnewline \cline{2-5}
				& $\sigma$					& 0.27 			& 0.39	& 0.20 \tabularnewline \hline

\multirow{2}{*}{ZSA}	& $\text{med}$ $(^{\circ})$		& 4.0 				& 6.1 		& 3.5 \tabularnewline \cline{2-5}
				& $\mu$ $(^{\circ})$			& 4.0				& 6.2		& 3.8 \tabularnewline \hline

\multirow{2}{*}{log(ZSA)} & $\mu$					& 0.6				& 0.72	& 0.55 \tabularnewline \cline{2-5}
				& $\sigma$ 					& 0.09			& 0.30	& 0.15 \tabularnewline \hline

\multirow{2}{*}{E[ZSA]}	& a 						& 0.05			& -0.002 	& -0.06 \tabularnewline \cline{2-5}
\multirow{2}{*}{ \big(See Eq.~(\ref{e6})   \big)} & b 						& -4.08			& 2.3		& 3.58 \tabularnewline \cline{2-5}
				& c 						& 0.81 			& 0.66	& 0.54 \tabularnewline \hline

\multirow{2}{*}{$K$-factor}	&$ \mu$				& 2.4				& -0.4		&1.5		\tabularnewline \cline{2-5}
					& $\sigma$				&2.0				& 4.3		&6.8		\tabularnewline \hline

Delay scaling			& $\mu$					& 3.9				&2.7		&3.2 		\tabularnewline \cline{2-5}
parameter $r_{DS}$		& $\sigma$				& 2.1				& 3.6		& 4.3 		\tabularnewline \hline

\end{tabular}
}
\label{tbl:T1}
\end{table}

\begin{table}[t]
\centering
\caption{Cross-correlation coefficients obtained from 28 GHz and 73 GHz ultrawideband channel measurements.}
\resizebox{8.5cm}{!} {
\begin{tabular}{|c|c|c|c|c|}
\hline

\multicolumn{2}{|c|}{\multirow{2}{*}{Scenario}} & \multicolumn{3}{c|}{Urban Microcell (UMi)} \tabularnewline \cline{3-5}
\multicolumn{2}{|c|}{}	& LOS		& \multicolumn{2}{c|}{NLOS}				\tabularnewline \cline{1-5}

\multicolumn{2}{|c|}{Parameters}	& 28-73 GHz Combined	& 28 GHz 	& 73 GHz \tabularnewline \hline

\multicolumn{2}{|c|}{ASD vs DS}					& 0.32	& -0.051	& 0.021	\tabularnewline \hline
\multicolumn{2}{|c|}{ASA vs DS}					& 0.49	& 0.153	& 0.264	\tabularnewline \hline
\multicolumn{2}{|c|}{ASA vs SF}					& 0.54	& -0.637	& 0.044	\tabularnewline \hline
\multicolumn{2}{|c|}{ASD vs SF}					& -0.04	& 0.051	& 0.008	\tabularnewline \hline
\multicolumn{2}{|c|}{DS vs SF}					& 0.35	& -0.508	& -0.187	\tabularnewline \hline
\multicolumn{2}{|c|}{ASD vs ASA}				& 0.72	& 0.405	& -0.257	\tabularnewline \hline
\multicolumn{2}{|c|}{ASD vs \textit{K}}				& -0.16	& -0.217	& 0.162	\tabularnewline \hline
\multicolumn{2}{|c|}{ASA vs \textit{K}}				& 0.07	& -0.069	& -0.428	\tabularnewline \hline
\multicolumn{2}{|c|}{DS vs \textit{K}}				& -0.46	& -0.133	& -0.449 	\tabularnewline \hline
\multicolumn{2}{|c|}{SF vs \textit{K}}				& -0.03	& -0.278	& 0.029	\tabularnewline \hline
\multicolumn{2}{|c|}{ZSA vs SF}					& 0.16	& -0.480	& -0.327	\tabularnewline \hline
\multicolumn{2}{|c|}{ZSA vs \textit{K}}				& -0.37	& -0.077	& -0.105	\tabularnewline \hline
\multicolumn{2}{|c|}{ZSA vs DS}					& 0.44	& 0.347	& 0.144	\tabularnewline \hline
\multicolumn{2}{|c|}{ZSA vs ASD}				& 0.95	& 0.042	& -0.027	\tabularnewline \hline
\multicolumn{2}{|c|}{ZSA vs ASA}				& 0.72	& 0.323	& 0.081	\tabularnewline \hline

\end{tabular}
}
\label{tbl:T2}
\end{table}

\section{Small-Scale Parameters}

The small-scale parameters specify the characteristics of individual propagation paths, and are used to generate path powers, path delays, AODs, and AOAs, and are commonly obtained using high-resolution joint delay-angle multipath parameter extraction algorithms, such as the SAGE~\cite{Fleury99} and KPowerMeans algorithms~\cite{Czink06}. The KPowerMeans algorithm provides an optimum assignment of multipath components into joint delay-angle clusters given a desired number of clusters from multi-dimensional channel impulse response data, using the mathematically-based power-weighted multipath component distance (MCD) metric~\cite{Czink06_2}. The optimum number of clusters is then determined from two optimal criteria, the Cali\~{n}ski-Harabasz and the Davies-Bouldin indices~\cite{Czink06_2}. The KPowerMeans algorithm was run 50 times to remove the effects of initialization of centroid starting positions. The parameters $t$, $s$, and $p$ were set to 2, 0.9, and 0.9, respectively, for the \textit{combineValidate} and \textit{shapePruning} steps. The average number of clusters and cluster subpaths were determined to be 5.0 and 12.4 for the combined 28-73 GHz LOS scenario, respectively, 5.3 and 12.8 for the 28 GHz NLOS environment, and 4.6 and 13.2 for the 73 GHz NLOS environments, as summarized in Table~\ref{tbl:T3}.

\begin{table}
\centering
\caption{Table summarizing small-scale model parameters, extracted using the KPowerMeans algorithm from 28 GHz and 73 GHz ultrawideband channel measurements.}
\resizebox{8.5cm}{!} {
\begin{tabular}{|c|c|c|c|}
\hline

\multirow{3}{*}{Scenario} & \multicolumn{3}{c|}{Urban Microcell (UMi)} \tabularnewline \cline{2-4} 
				&	LOS		& \multicolumn{2}{c|}{NLOS} \tabularnewline \cline{2-4}
				& 	Combined 28-73 GHz 	& 28 GHz & 73 GHz \tabularnewline \hline
Clustering	& \multicolumn{3}{c|}{KPowerMeans} \tabularnewline \hline

\# of clusters $(\mu,\sigma)$ 		& 5.0, 3.5		& 5.3, 2.4 	& 4.6, 3.3 \tabularnewline \hline

\# of subpaths $(\mu,\sigma)$ 		& 12.4, 14.8		& 12.8, 21.4	& 13.2, 22.1 \tabularnewline \hline

Cluster ASD ($^{\circ}$) $(\mu,\sigma)$ 	& 1.5, 2.2		& 3.0, 4.5	& 2.1, 6.9	\tabularnewline \hline

Cluster ASA ($^{\circ}$)	 $(\mu,\sigma)$ 	& 6.7, 16.1		& 9.6, 20.1	& 5.2, 12.1	\tabularnewline \hline

Cluster ZSD ($^{\circ}$)	$(\mu,\sigma)$ 	& 0.8, 1.0			& - 		& 0.8, 1.15	\tabularnewline \hline

Cluster ZSA ($^{\circ}$)	$(\mu,\sigma)$ 	& 1.8, 2.0		& 1.6, 3.5	& 1.5, 1.9  \tabularnewline \hline

Per-cluster	& \multirow{2}{*}{13.6}	&	 \multirow{2}{*}{16.1} &  \multirow{2}{*}{17.4}  \tabularnewline
shadowing (dB) &				&					& \tabularnewline \hline

\end{tabular}
}
\label{tbl:T3}

\end{table}

A flaw of the KPowerMeans algorithm is that it fails to converge to one final solution as a result of the random initialization procedure, that assigns mutipath components to random delay-angle clusters. Every time the algorithm is run, the first assignment of multipaths to clusters is arbitrary, that can lead to differences in the final cluster partition. In addition, the Cali\~{n}ski-Harabasz index $CH(K)$ carries an undefined numerator for a value of $K=$ 1 cluster, as shown below~\cite{Czink06_2}: 
\begin{equation}\label{eqCH}
CH(K) = \frac{\text{tr}(\bm{B}) / (K-1)}{\text{tr}(\bm{W}) / (L-K)}
\end{equation}
\noindent where,
\begin{align}
&\text{tr} (\bm{B}) = \sum_{k=1}^{K} L_k MCD(\bm{c_k},\bm{\overline{c}})^2 \\
&\text{tr} (\bm{W}) =\sum_{k=1}^{K}\sum_{j \in C_k} MCD(x_j, \bm{\overline{c}}_k)^2
\end{align} 
\noindent where $K$ is an integer representing the desired number of clusters, $L$ is the total number of paths considered, $L_k$ is the number of paths in the $k$\textsuperscript{th} cluster, $MCD(x_i,x_j)$ is the multipath component distance between vectors $x_i$ and $x_j$, $\bm{\overline{c}}$ is the global centroid of the data, and $\bm{\overline{c}}_k$ is the centroid of the $k$\textsuperscript{th} cluster. It is clear that~(\ref{eqCH}) is undefined for $K=1$, thus suggesting that $K=1$ cluster can never be optimum for an arbitrary dataset. The fine-tuning of parameters $t$, $s$, $p$, and the number of total runs of the algorithm are not explicitly discussed and remain open issues in~\cite{Czink06_2}. In addition, the \textit{shapePruning} step in~\cite{Czink06_2} discards outlier data that do not significantly affect the properties of final clusters.

The \textit{time cluster - spatial lobe} (TCSL) algorithm~\cite{Samimi15_3} offers an alternative \textit{physically}-based binning procedure, where a \textit{time cluster} corresponds to a group of traveling multipaths with similar delays but with potentially varying AOAs, and a \textit{spatial lobe} denotes a strong direction of arrival (or departure) where energy is received contiguously in the azimuth and/or elevation dimensions. Statistics of time clusters are obtained using a physically-based 25 ns minimum inter-cluster void interval~\cite{Samimi15_3} in the time domain (representing 8 m in propagation distance, typical minimum spatial voids between narrow streets or buildings in New York City) by partitioning omnidirectional PDPs based on time of arrivals, since multipaths tend to separate due to free space gaps in the environment between buildings and other larger reflectors. Separately, the characteristics of spatial lobes are extracted by defining a -10 dB or -20 dB lobe power threshold with respect to the maximum peak received power in the 3-D power angular spectrum, where all contiguous angular segments above the lobe power threshold constitute one 3-D spatial lobe. After performing the separate time and spatial domain clustering, by randomly allocating the time delay multipaths to particular spatial lobes, the first- and second-order temporal and spatial statistics are well reproduced~\cite{Samimi15_2,Samimi15_3}. Table~\ref{tbl:T23} presents key parameters describing time clusters and spatial lobes obtained from all available omnidirectional PDPs and TX/RX angular spectra, used to generate mmWave channel coefficients~\cite{Samimi15_3}.

\begin{table*}
\centering
\captionsetup{width=\textwidth}
\caption{Key parameters describing time clusters and spatial lobes to generate mmWave channel coefficients~\cite{Samimi15_3,Samimi15_2}. A `-' indicates that the data is unavailable from the measurements, which considered one omnidirectional azimuth scan for one fixed TX elevation downtilt at 28 GHz (LOS and NLOS), and 73 GHz LOS~\cite{Rap15_3}.}
\begin{tabular}{|c|c|c|c|c|c|c|}
\hline
\multicolumn{2}{|c|}{\multirow{2}{*}{Scenario}} & \multicolumn{5}{c|}{Urban Microcell (UMi)} \tabularnewline \cline{3-7}
\multicolumn{2}{|c|}{}								& \multicolumn{3}{c|}{LOS} &	\multicolumn{2}{c|}{NLOS} \tabularnewline \hline

\multicolumn{2}{|c|}{Parameters}	& 28 GHz	& 73 GHz	& Combined 28-73 GHz	& 28 GHz	& 73 GHz \tabularnewline \hline

Number of clusters	& $(\mu,\sigma)$	& 3.0, 2.0	& 1.8, 0.8	& 2.3, 1.4 	& 2.1, 1.4 & 2.7, 1.4 \tabularnewline \hline

Number of subpaths & $(\mu,\sigma)$	& 7.4, 5.8	& 7.8, 5.0	& 7.6, 5.2	&9.1, 10.9 & 5.7, 6.0 \tabularnewline \hline

Cluster decay constant	& $\Gamma$ (ns)	& 38.6 & 17.5	& 25.9	&49.4	& 56.0	\tabularnewline \hline

Per-cluster shadowing 	& $\sigma$ (dB)	& \multicolumn{3}{c|}{1}		& \multicolumn{2}{c|}{3} \tabularnewline \hline

Subpath decay constant	& $\gamma$ (ns)	& 25.2	& 13.0	& 16.9	& 16.9 & 15.3 \tabularnewline \hline

Per-subpath shadowing	& $\sigma$ (dB)	& \multicolumn{5}{c|}{6} \tabularnewline \hline \hline

\# of AOD spatial lobes	& $(\mu,\sigma)$	& 3.3, 0.6	& 1.0, 0	&1.9, 1.2	& 1.6, 1.8	& 1.5, 0.7 \tabularnewline \hline

\# of AOA spatial lobes	& $(\mu,\sigma)$	& 2.3, 1.5	& 1.4, 0.5	& 1.8, 1.0	& 1.6, 0.7	& 2.5, 1.1 \tabularnewline \hline

AOD/AOA azimuth angles & \multicolumn{6}{c|}{Uniform(0,360)} \tabularnewline \hline

AOD/AOA elevation angles &	\multicolumn{6}{c|}{Gaussian (AOD) / Laplacian (AOA)} \tabularnewline \hline

RMS lobe ASD		& $(\mu,\sigma)$ (deg)	& 6.0, 3.2	& 4.7, 1.0	& 5.6, 2.7	&6.2, 3.3	& 4.9, 2.7 \tabularnewline \hline

RMS lobe ESD		& $(\mu,\sigma)$ (deg)	& - 		& - 		& - 	& -& 2.2, 0.8 \tabularnewline \hline

RMS lobe ASA		& $(\mu,\sigma)$ (deg)	& 10.1, 3.9	& 3.9, 0.8	& 7.0, 4.2	& 6.8, 4.8	& 3.7, 2.3 \tabularnewline \hline

RMS lobe ESA		& $(\mu,\sigma)$ (deg)	& 11.2, 2.7 &	2.9, 0.7	& 7.0, 4.7	& 6.7, 2.3 &	2.2, 1.7 \tabularnewline \hline

\end{tabular}
\label{tbl:T23}

\end{table*}

\section{Number of Multipath Components for Directional Pointing Angles}

Table~\ref{tbl:T11} shows the number of multipath components measured at unique TX-RX directional pointing angles at 28 GHz and 73 GHz, obtained using a peak detection algorithm. The mean number of multipath components per unique pointing angle was obtained by considering all directional PDPs with at least 5 dB signal-to-noise (SNR) ratio, and then averaging the number of resolvable multipaths in all profiles over all RX locations. The mean number of multipath components per unique pointing angle was 7.2 and 3.8 at 28 GHz and 73 GHz in LOS, respectively, and 5.2 and 3.3 at 28 GHz and 73 GHz in NLOS, respectively. The mean number of multipath components at 28 GHz is larger than at 73 GHz in LOS environments, and may result from the larger beamwidth used at 28 GHz (10.9$^{\circ}$ beamwidth), and from the difference in measurement procedures, where the 28 GHz measurements used rigid TX and RX pre-determined angle combinations, whereas the 73 GHz angles were determined based on strongest received power from exhaustive initial searching during the field measurements. Since the 28 GHz measurements did not always consider strongest elevation planes, diffuse scattering was found to be more prominent at the weaker measured angles at 28 GHz, carrying many more multipath components than for the 73 GHz measurements, where fewer, but stronger multipath components were detected. Note that in~\cite{Rap15_3}, a different methodology was used to compute the mean number of multipath components at unique pointing angles: the numbers of detected multipath components in all directional PDPs satisfying at least a 5 dB SNR were first averaged over all pointing angles at each RX location, and then averaged again over all RX locations, yielding means of 4.7 and 3.3 at 28 GHz and 73 GHz in NLOS, respectively.

\begin{table}
\centering
\caption{Mean and standard deviation of the number of multipath components for directional TX-RX arbitrary pointing angles~\cite{Rap15_3}.}
\begin{tabular}{|c|c|c|}
\hline
\multicolumn{3}{|c|}{Number of Multipaths (Directional)} \tabularnewline \hline
Scenario 			& LOS	& NLOS \tabularnewline \hline
28 GHz $(\mu, \sigma)$	& 7.2, 5.3	& 5.2, 4.6 \tabularnewline \hline
73 GHz $(\mu, \sigma)$	& 3.8, 3.1	& 3.3, 2.7 \tabularnewline \hline

\end{tabular}
\label{tbl:T11}
\end{table}

\section{Cross-Polar Ratio (XPR) Measurements}

\begin{figure}[t]
    \centering
 \includegraphics[width=3.5in]{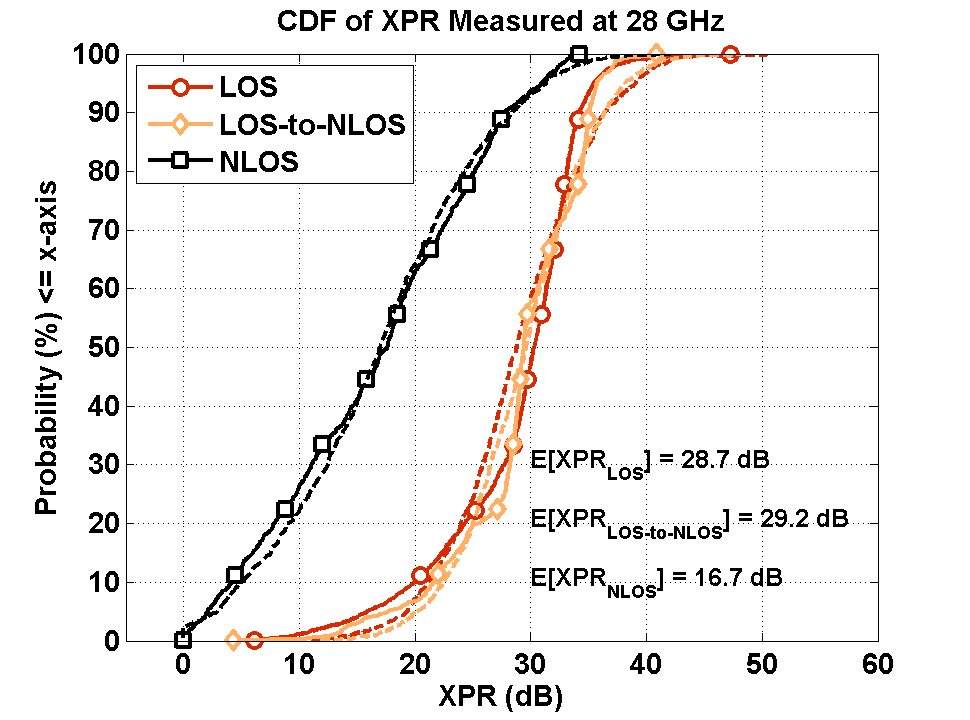}
    \caption{Cumulative distribution function of 28 GHz ultrawideband cross-polarization ratio (XPR) measurements for LOS, LOS-to-NLOS, and NLOS environments, and corresponding simulated Gaussian distributions.}
    \label{fig:xpr}
\end{figure}

28 GHz XPR small-scale fading measurements investigated the effects of polarization on resolvable multipath components by sampling PDPs at half-wavelength spatial increments over a 33-wavelength long linear track.  For each 2.5 ns time delay bin, the XPR was extracted by dividing the received power $P_{VV}$ (mW) from vertically-polarized TX and RX antennas, by the received power $P_{VH}$ (mW) from vertically-polarized TX and horizontally-polarized RX antennas. Fig.~\ref{fig:xpr} shows the cumulative distribution functions (CDFs) for the measured XPR in LOS, LOS-to-NLOS, and NLOS environments, and simulated Gaussian CDFs with values in dB. The mean and standard deviation of the XPR are summarized in Table~\ref{tbl:T4}, and were found to be 28.7 dB and 6.0 dB in LOS, 29.2 dB and 5.5 dB in LOS-to-NLOS, and 16.7 dB and 8.8 dB in NLOS, respectively. The XPR curves closely follow a Gaussian distribution (in dB), with the following probability density function (PDF):
\begin{equation}
f(x) = \frac{1}{\sqrt{2 \pi} \sigma}e^{-\frac{(x - \mu)^2}{2\sigma^2}}
\end{equation}

\noindent where $\mu$ and $\sigma$ are set to the mean and standard deviation (in dB) of the measurement set. Note that in Fig~\ref{fig:xpr}, the simulated data is obtained by taking the maximum with respect to 0 dB. In NLOS, the mean XPR is 12 dB smaller than in LOS, indicating that the propagating radio-waves experience depolarization from reflections and scattering in the environment. 

\begin{table}
\centering
\caption{Means ($\mu$) and standard deviations ($\sigma$) of the XPR obtained from small-scale fading track measurements at 28 GHz.}

\begin{tabular}{|c|c|c|c|c|}

\hline
\multicolumn{2}{|c|}{\multirow{2}{*}{Scenario}} & \multicolumn{3}{c|}{Outdoor Street Canyon} \tabularnewline \cline{3-5} 
\multicolumn{2}{|c|}{}	& LOS	& LOS-to-NLOS	& NLOS \tabularnewline \hline

\multirow{2}{*}{XPR (dB)} & $\mu$ 	& 28.7 	& 29.2	& 16.7 \tabularnewline \cline{2-5}

				& $\sigma$ 	& 6.0 	& 5.5 	& 8.8 \tabularnewline \hline
\multicolumn{2}{|c|}{Truncated Distribution} & \multicolumn{3}{c|}{$\max \big\{   N(\mu,\sigma^2),0  \big\}$}  \tabularnewline \hline

\end{tabular}
\label{tbl:T4}

\end{table}

\section{Conclusion}

This paper presented 28 GHz and 73 GHz large-scale and small-scale model parameters required to generate the temporal and angular characteristics of multipath components in a PDP. The RMS DS and SF were found to be negatively correlated at both 28 GHz and 73 GHz, with a value of \text{-0.508} and \text{-0.187} in NLOS, respectively, while the ASA and SF were found to be negatively correlated at 28 GHz in NLOS with a value of \text{-0.637} and nearly uncorrelated at 73 GHz in NLOS with a value of 0.044. Joint delay-angle clustering approaches, such as the KPowerMeans algorithm, use a mathematically-based clustering scheme to extract cluster statistics. Instead, the TCSL algorithm employs a physically-based thresholding technique separately in delay and angle to extract multipath properties~\cite{Samimi15_3}. The statistics presented here can be used in realistic system-level simulations and air-interface design~\cite{Fung93} of next generation mmWave communication systems.

\vspace{7pt}
\bibliographystyle{IEEEtran}
\bibliography{bibliography_MSThesis}
\end{document}